\documentclass[journal]{IEEEtran}
\usepackage[dvipdfmx]{graphicx}
\usepackage{booktabs}
\usepackage{pifont}
\usepackage{float}      
\usepackage{comment}
\usepackage{color}
\usepackage{amsmath}
\usepackage{url}
\usepackage{algorithm}
\usepackage{algorithmic}
\usepackage{amsmath}
\usepackage{bbm}

\usepackage{amsfonts}
\usepackage{bm}

\ifCLASSINFOpdf
\else
\fi

\begin{document}

\title{Query Timing Analysis for Content-based Wake-up Realizing Informative IoT Data Collection}
\author{Junya~Shiraishi,~\IEEEmembership{Graduate Student Member,~IEEE,}
        Anders~E.~Kal\o r,~\IEEEmembership{Graduate Student Member,~IEEE,}
        Federico~Chiariotti,~\IEEEmembership{Member,~IEEE,}
        Israel~Leyva-Mayorga,~\IEEEmembership{Member,~IEEE,}
        Petar~Popovski,~\IEEEmembership{Fellow,~IEEE,}
        and~Hiroyuki~Yomo,~\IEEEmembership{Member,~IEEE}\thanks{J. Shiraishi and and H. Yomo are with the Graduate School of Science and Engineering, Kansai University, Japan (email: $\{$k980158, yomo$\}$@kansai-u.ac.jp). A. E. Kal{\o}r, F. Chiariotti, I. Leyva-Mayorga, and P. Popovski are with the Department of Electronic Systems, Aalborg University, Denmark (email: $\{$aek, fchi, ilm, petarp$\}$@es.aau.dk). The work of J. Shiraishi and H. Yomo was supported by JSPS KAKENHI under Grant JP22K04114, by JST SPRING, Grant Number JPMJSP2150, and by JSPS Overseas Challenge Program for Young Researchers. The work of A. E. Kal{\o}r was supported by the Independent Research Fund Denmark (IRFD) under Grant 1056-00006B. The work of F. Chiariotti, I. Leyva-Mayorga and P. Popovski was supported by the Villum Investigator Grant ``WATER'' from the Velux Foundation, Denmark. }}%

\maketitle

\begin{abstract}
Information freshness and high energy-efficiency are key requirements for sensor nodes serving Industrial Internet of Things (IIoT) applications, where a sink node must collect informative data before a deadline to control an external element. Pull-based communication is an interesting approach for optimizing information freshness and saving wasteful energy. To this end, we apply Content-based Wake-up (CoWu), in which the sink can activate a subset of nodes observing informative data at the time that wake-up signal is received. In this case, the timing of the wake-up signal plays an important role: early transmission leads to high reliability in data collection, but the received data may become obsolete by the deadline, while later transmission ensures a higher timeliness of the sensed data, but some nodes might not manage to communicate their data before the deadline. This letter investigates the timing for data collection using CoWu and characterizes the gain of CoWu. The obtained numerical results show that CoWu improves accuracy, while reducing energy consumption by about $75\%$ with respect to round-robin scheduling.
\end{abstract}


\IEEEpeerreviewmaketitle

\section{Introduction}
\IEEEPARstart{L}{ow-power} wireless sensor networks (WSNs) are expected to play a key role in supporting novel Industrial Internet of Things (IIoT) \cite{sisinni2018industrial} applications, where the collected data may be used for real-time monitoring and diagnostics, and to control actuators. To support these applications, it is crucial for the sensor data to not only be available for the monitoring/control process when needed, but also be \emph{timely}, i.e., represent the current state of the system. However, ensuring the timeliness of the data requires frequent transmissions, which in turn increases the power consumption of the nodes. An attractive strategy to ensure both timeliness and low power consumption is to operate in the \emph{pull-based} communication regime, in which the desired data is directly requested by a sink node prior to a deadline, as opposed to being transmitted regularly by the sensors~\cite{chiariotti2022query}. In addition to ensuring that the data is transmitted only when it is needed, this also allows the sink to request specific types of data, such as values within a certain range, which further reduces the total number of sensor transmissions and thus the power consumption of the WSN.

\begin{figure}[t]
\centering
\includegraphics[width=0.40\textwidth]{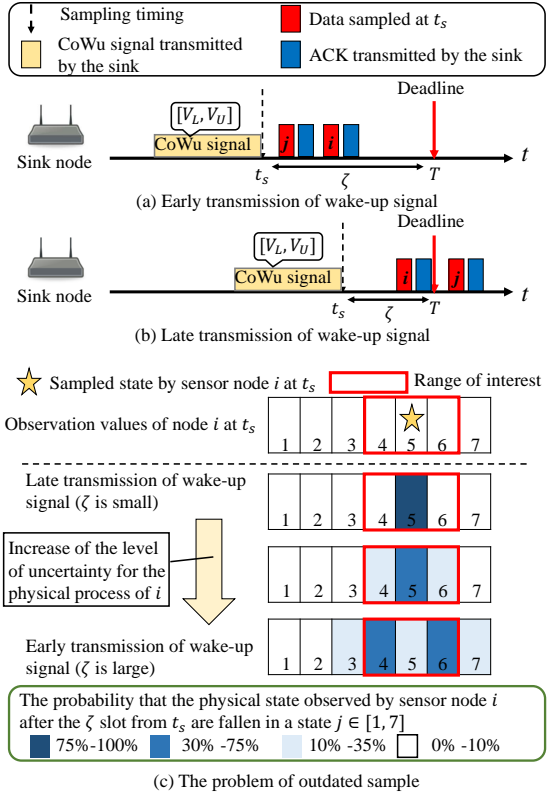}
\caption{An example of data collection employing CoWu.}
\label{Fig:Data_Collection_Model_CoWu}
\end{figure}

In this letter, we propose a pull-based, timely communication scheme using wake-up radio~\cite{IEICE-yomo,piyare2017ultra}. By employing wake-up receivers at the sensors, a sensor's primary radio can be turned off when there is no need to communicate, while an ultra-low power wake-up receiver is always listening for a wake-up request from the sink node. The wake-up radio can be triggered either using identity-based wake-up (IDWu)~\cite{WakeupR} or Content-based Wake-up (CoWu)~\cite{TGCN_Content}. In IDWu, the wake-up condition is based on the ID of each node, thereby allowing the sink to target a specific sensor. On the other hand, in CoWu the wake-up condition is based on the data that is currently being observed by a given node. We focus on the case where the sink aims to collect sensor values within a certain range. To this end, we apply a CoWu strategy and characterize the timing of the wake-up signal: sending it too early increases the risk that the values will be outdated, i.e., no longer within the requested range, at the deadline, while sending it too late may prevent some of the sensors from successfully transmitting their values before the deadline. This trade-off is illustrated in Fig.~\ref{Fig:Data_Collection_Model_CoWu}(a) and \ref{Fig:Data_Collection_Model_CoWu}(b), where the range $[V_L,V_U]$ is requested from the sensors $\zeta$ time slots prior to the deadline, and the sink transmits the wake-up signal in an early and late timing, respectively. Similarly, Fig. \ref{Fig:Data_Collection_Model_CoWu}(c), illustrates an example of how $\zeta$ influences the likelihood that an observed process taking values in the interval $[1,7]$, remains in the requested range $[4, 6]$ from the request time, $t_s$, until the deadline, $T$. We present a theoretical analysis considering a realistic medium access control (MAC) protocol, evaluate the performance of CoWu with the baseline scheme, and clarify the importance of the timing of the CoWu signaling and the performance of CoWu as a function of the speed of the physical process. 

The importance of timely information in communication system has been studied extensively in the literature on the Age of Information (AoI) metric~\cite{kaul2011minimizing,kosta2017age}, which measures the time elapsed since the generation of the last measurement received by the sink node. The AoI of pull-based transmission strategies has previously been studied in~\cite{chiariotti2022query}, and the related Value of Information (VoI) metric was analyzed in~\cite{chiariotti2022scheduling}. The use of wake-up radio to collect data has been studied, amongst others, in the context of Unmanned Aerial Vehicles (UAVs) \cite{trotta2019bee}, and to obtain the top-$k$ values in a WSN~\cite{shiraishi2021periodical}. To the best of our knowledge, this is the first work that characterizes the effect of timing issues in wake-up radio, particularly when the data is needed at a specific time. Our contributions are twofold. First, we analyze the impact of the wake-up signal timing with respect to a deadline, considering the evolution of the physical process. Second, we clarify the gain of CoWu and investigate its robustness against the estimation error.

The remainder of the letter is divided as follows. In Sec.~\ref{sec:sys} we present the system model for the observed process and the wake-up signal model. We analyze the performance of the considered scheme in Sec.~\ref{sec:Theoretical_Analysis}, and present our results in Sec.~\ref{sec:sim}. Finally, we conclude the letter in Sec.~\ref{sec:conc} and discuss possible avenues for future work.

\section{System Model and Problem Definition}\label{sec:sys}
\subsection{Scenario and Objective}
We study a time-slotted scenario comprising a sink node and $N$ sensor nodes equipped with wake-up receivers. Each sensor observes an independent but identical, integer-valued physical process that takes values $1,2,\ldots, M$, and evolves according to an $M$-state discrete-time, irreducible Markov process with transition matrix $\mathbf{Z}$. 
The time is indexed by the time slots, and we will denote the steady-state distribution of the Markov process by $\bm{\pi} = \{\pi_{1},\pi_{2}, \ldots, \pi_{M}\}$. The process dynamics, but not the instantaneous values, are assumed to be known by the sink node.

The sink node receives sporadic requests, i.e., at unpredictable time instants, to collect data from the sensor nodes, which must be gathered before a given deadline $T$. The requests originate from an external entity, such as an actuator or monitoring software. To facilitate our analysis, in this initial work, we will ignore the time dependency between requests, i.e., we will assume that the time between consecutive requests is sufficiently long to allow the physical processes to reach steady-state conditions. 
This assumption is reasonable in many applications, e.g., in the mobile actuators scenario \cite{zeng2013real}, in which the query from an actuator arrives only when the mobile actuator enters the area of interest. 
We limit our focus to \emph{range queries}, i.e., where the actuator requests sensor readings whose values are in the interval $[V_{L}, V_{U}]$ where $1\le V_{L}\le V_{U}\le M$. 

The data is collected by the sink node by forwarding the range query to the sensors using CoWu at time $t_s$. In this letter, we assume that each node samples its data simultaneously in CoWu. The analysis of different sampling policies considering their practical feasibility is kept for future work. 
Since the physical processes generally evolve between the sampling time and the deadline, we define $\zeta=T-t_s$ as the time interval between the sampling time and the deadline, measured in time slots. The goal of the sink is to collect data that \emph{remains timely} at the deadline. To this end, we define the accuracy, $\gamma$, as the probability that the received sensor measurements are exactly the ones that are within the queried range at the time of the deadline. Formally, denoting by $\mathcal{T}$ the subset of nodes belonging to the interval [$V_{L}, V_{U}$] at the deadline time $T$, and by $\mathcal{S}$ the subset of nodes whose data has been successfully transmitted to the sink by time $T$, we define the accuracy as
\begin{equation}
    \gamma = \Pr(\mathcal{T}=\mathcal{S}).\label{eq:gamma_Def}
\end{equation}
Thus, the accuracy is one only if the received range set always coincides with the true set at the deadline. Our aim is to characterize the trade-off between $\zeta$ and $\gamma$. The above model can be applied, e.g., for anomaly detection and its remedy in industrial applications or environmental monitoring and control, in which the sink conducts a range query to detect extreme values of the machine state, pollution level, etc. In this scenario, timely data collection is important to take an appropriate and prompt action.

\subsection{CoWu Transmission Model}\label{sec:sysmodel_cowu}
In CoWu~\cite{TGCN_Content}, the wake-up condition, such as the range interval, is embedded into a wake-up signal which is transmitted to the wake-up receivers through a secondary, low-power radio link, which applies a simple communication method, such as On-Off Keying (OOK). During this operation, only the wake-up receiver is active, while the primary radio is switched off, allowing the power consumption of the sensor node to be as low as a few microwatts, which is much smaller than the primary radio's. As a specific example, CoWu for the range query can be implemented by encoding the lower and upper interval limits, $V_L$ and $V_U$, into the duration of the wake-up signal \cite{TGCN_Content}. Specifically, the sink transmits first a wake-up signal of length proportional to the lower interval limit, $V_L$, and then transmits one proportional to the upper limit, $V_{U}$. Each wake-up receiver then checks whether its sensed value is within the range based on the signal length extracted by non-coherent OOK detection, and, if so, activates its main radio interface and transmits its observation; otherwise, it remains in a sleep state, keeping the main radio off.

The primary radio, used by the nodes that observe data within the specified range to transmit their observations to the sink, follows a $p$-persistent Carrier Sense Multiple Access (CSMA) protocol \cite{TGCN_Content}. We assume that each transmission occupies $L$ slots, and that the sink transmits an error-free acknowledgment after each successful sensor transmission, so that nodes return to sleep after a successful transmission. The $p$-persistent model, in which each active node transmits in an idle slot with probability $p$ and stays silent with probability $1-p$, is an analytically tractable, yet good approximation, of many practical CSMA protocols, such as the one used in the IEEE 802.15.4 standard~\cite{TGCN_Content,IEEE15.4,epoch}. For simplicity, we assume that all nodes, including the sink, are located within each other's communication/wake-up/carrier-sensing range, i.e., there are no hidden terminals.

\section{Analysis of CoWu}\label{sec:Theoretical_Analysis}
In this section, we derive the accuracy, $\gamma$, of the range query defined in Eq. (\ref{eq:gamma_Def}) for a given value of $\zeta$. In CoWu, the probability of a node waking up depends on $N$, the distribution of physical process, and $[V_{L}, V_{U}]$. 
Let $P_{w}(V_{th})$ denote the probability of nodes waking up given the stationary distribution, $\bm{\pi}$, and the CoWu threshold, $V_{th}=[V_{L}, V_{U}]$. We then have
\begin{equation}
P_{w}(V_{th})=\sum_{i=V_{L}}^{V_{U}}{\pi_{i}},
\end{equation}
and the probability that $w$ nodes wake up follows a binomial distribution
\begin{equation}
\begin{split}{
\ P_{d}(w)=\binom{N}{w}P_{w}(V_{th})^{w}(1-P_{w}(V_{th}))^{N-w}.\label{eq:distribution_nwake}
}\end{split}
\end{equation}

To compute the distribution of the number of successful transmissions under $p$-persistent CSMA, we construct a two-dimensional Markov chain indexed by the current slot, in which the state represents the number of nodes that still need to transmit and the number of slots elapsed since the start of the ongoing transmission. Specifically, conditioned on the number of active nodes $w$ and knowing that each transmission requires $L$ slots, the state space is $\{(w,0), (w, 1), \ldots, (w, L-1), (w-1, 0), \ldots, (1, L-1), (0,0)\}$, where state $(n, l)$ represents the case where $n$ nodes have not completed their transmission, and the transmitting node(s) has been transmitting for $l$ slots.

The transition probabilities are defined as follows. When the state is $(n, 0)$, $n=1,2,\ldots,w$, the channel is idle and one or more of the $n$ remaining nodes can initiate a transmission, causing a transition to state $(n,1)$. This happens with probability $1-(1-p)^n$. On the other hand, if none of the nodes transmit, the Markov chain remains in state $(n,0)$, which happens with probability $(1-p)^n$.  In states $(n, l)$, $l = 1, 2, \ldots, L-2$, the channel is busy and $L-l$ slots remain of the current transmission, so the Markov chain transitions to state $(n, l+1)$ with a probability 1. In state $(n, L-1)$, there are two cases depending on whether one or multiple users were transmitting in the previous slots. If only one node transmitted, the transmission is successful and the Markov chain transitions into state $(n-1, 0)$. This happens with probability
\begin{equation}
S_{n}=\frac{np(1-p)^{n-1}}{1-(1-p)^n},
\end{equation}
which is conditioned on the event that at least one node is transmitting. If more than one user transmitted, which happens with probability $1-S_{n}$, all transmissions fail and the Markov chain returns to state $(n, 0)$. Finally, state $(0, 0)$ is an absorbing state representing the event that all $w$ active users have successfully transmitted their measurement.

Using the Markov chain, we can obtain the distribution of the number of successful transmissions by state evolution from the initial state distribution $\Phi(0) = (1, 0, 0, \ldots, 0)$ as
\begin{equation}
\Phi(t+1)=\Phi(t){\mathbf{R}},
\end{equation} 
where $\mathbf{R}$ is $(wL+1) \times (wL+1)$ transition matrix containing the transition probabilities defined above and $\Phi(t)\in[0,1]^{(wL +1)}$ is the state vector representing the probability of each state at time $t$, whose entry corresponding to state $(i, j)$ is denoted as $\phi_{(i, j)}(t)$. The probability that $w_{s}$ out of the $w$ active nodes succeed by the deadline for a given $\zeta$, denoted as $P_{s}(w_{s}, \zeta)$, is then
\begin{equation}\begin{split}
P_{s}(w_{s}, \zeta)&=\begin{cases}\phi_{(0,0)}(\zeta)&\text{if}~ w_{s}=w\\\sum_{l=0}^{L-1}\phi_{(w-w_{s},l)} (\zeta) & \text{otherwise.}\label{eq:n_s_D}
\end{cases}\end{split}\end{equation}

We can now derive the accuracy, defined as the probability that the set of nodes from which the sink has received values coincides with the set of nodes whose reading at the deadline, $T$, is in the requested interval $[V_{L}, V_{U}]$. Let us denote the state of the physical process at node $i$ at time $t_{s}$ as $v_{s}^{i}$ and at the deadline $T$ as $v_{T}^{i}$.
In order for the sink to estimate the range set correctly at the deadline $T$, all of the following three conditions must be satisfied:
\begin{itemize}
    \item{Cond. A: For the nodes that succeed in data transmission, $(v_{s}^{i} \in [V_{L}, V_{U}]) \land (v_{T}^{i} \in [V_{L}, V_{U}])$}.
    \item{Cond. B: For the nodes that wake up but fail their data transmission by $T$, $(v_{s}^{i} \in [V_{L}, V_{U}]) \land (v_{T}^{i} \notin [V_{L}, V_{U}])$}.
    \item{Cond. C: For the nodes that do not wake up and transmit data, $(v_{s}^{i} \notin [V_{L}, V_{U}]) \land (v_{T}^{i} \notin [V_{L}, V_{U}])$}.
\end{itemize}
Due to the symmetry of the physical processes, the probabilities of conditions A, B, and C are the same for all users, and denoted as $P_{A} (\zeta)$, $P_{B} (\zeta)$ and $P_{C} (\zeta)$, respectively. The probabilities are given as 
\begin{equation}
\ P_{A}(\zeta)=\frac{\sum_{i \in [V_{L}, V_{U}]}\left[\pi_{i}\sum_{j \in [V_{L}, V_{U}]}{[\mathbf{Z}^{\zeta}]_{i,j}}\right]}{\sum_{s \in [V_{L}, V_{U}]}\pi_{s}},\label{eq.P_A}
\end{equation}
\begin{equation}
\ P_{B}(\zeta)=\frac{\sum_{i \in [V_{L}, V_{U}]}\left[\pi_{i}\sum_{j \notin [V_{L}, V_{U}]}{[\mathbf{Z}^{\zeta}]_{i,j}}\right]}{\sum_{s \in [V_{L}, V_{U}]}\pi_{s}},\label{eq.P_B}
\end{equation}
\begin{equation}
\ P_{C}(\zeta)=\frac{\sum_{i \notin [V_{L}, V_{U}]}\left[\pi_{i}\sum_{j \notin [V_{L}, V_{U}]}{[\mathbf{Z}^{\zeta}]_{i,j}}\right]}{\sum_{s \notin [V_{L}, V_{U}]}\pi_{s}},\label{eq.P_C}
\end{equation}
where $[\mathbf{Z}]_{i,j}$ is the $(i,j)$-th entry of $\mathbf{Z}$.
Combining Eqs. (\ref{eq:distribution_nwake}), (\ref{eq:n_s_D}), (\ref{eq.P_A}), (\ref{eq.P_B}), and (\ref{eq.P_C}), the accuracy of CoWu for a given $\zeta$ and $[V_{L}, V_{U}]$ can be computed as
\begin{equation}
\begin{split}
\gamma_{CoWu}(\zeta) = \sum_{w=0}^{N}\sum_{w_{s}=0}^{w}&{P_{A}(\zeta)}^{w_{s}}{P_{B}(\zeta)}^{w-w_{s}}\\
&\times {P_{C}(\zeta)}^{N-w}P_{s}(w_{s}, \zeta)P_{d}(w).
\end{split}
\end{equation}

\section{Numerical results}\label{sec:sim}
In this section, we evaluate the performance of CoWu and compare it to a round-robin scheduling method. Throughout the evaluation, we assume that the physical process follows a truncated birth-death process where the probability that the value is incremented or decremented is $q$. 

\subsection{Baseline scheme for evaluation: Round-Robin Scheduling}\label{sec:sysmodel_rr}
In round-robin scheduling, the nodes transmit their measurements according to a Time Division Multiple Access (TDMA)-like policy, whose transmission starts exactly $NL$ slots prior to the deadline, denoted as $t_{sch}$, after detecting a wake-up signal triggering all nodes at each wake-up receiver.  
Then, the node $j=0, 1, \ldots, N-1$, samples and transmits its measurement at $t_{sch}+jL$.

Here, we derive the accuracy $\gamma$ of round-robin scheduling for the range query. 
The distribution of physical process during the sampling period is assumed to be the stationary distribution, and sampled value evolves by the discrete-Markov chain model until the deadline. 
The value to be taken at the deadline $T$ to estimate the correct range-set at the sink depends on the value observed by each node in each slot. In round-robin scheduling, accuracy is perfect only if all nodes satisfy the following requirements: 
\begin{itemize}
\item {Cond. D: If $v_{s}^{i} \in [V_{L}, V_{U}]$, then $v_{T}^{i} \in [V_{L}, V_{U}]$}
\item {Cond. E: If $v_{s}^{i} \notin [V_{L}, V_{U}]$, then $v_{T}^{i} \notin [V_{L}, V_{U}]$.}
\end{itemize}
Here, the probabilities of conditions D and E are denoted as $P_{D} (\zeta)$ and $P_{E} (\zeta)$, respectively, and can be expressed as
\begin{equation}
P_{D}(\zeta)=\sum_{i \in [V_{L}, V_{U}]}\left[\pi_{i}\sum_{j \in [V_{L}, V_{U}]}{[\mathbf{Z}^{\zeta}]_{i,j}}\right],\label{eq:P_D}
\end{equation}
\begin{equation}
P_{E}(\zeta)=\sum_{i \notin [V_{L}, V_{U}]}\left[\pi_{i}\sum_{j \notin [V_{L}, V_{U}]}{[\mathbf{Z}^{\zeta}]_{i,j}}\right].\label{eq:P_E}
\end{equation}
The accuracy of round-robin scheduling is then computed as
\begin{equation}
\gamma_{Sch}=\prod_{i=1}^{N}P_{D}\{(N-i+1)L\}+P_{E}\{(N-i+1)L\}.
\end{equation}
\subsection{Comparison between CoWu and round-robin scheduling}
\subsubsection{Accuracy against $\zeta$}\label{sec:D_opt}
\begin{figure}[t]
\centering
\includegraphics[width=0.48\textwidth]{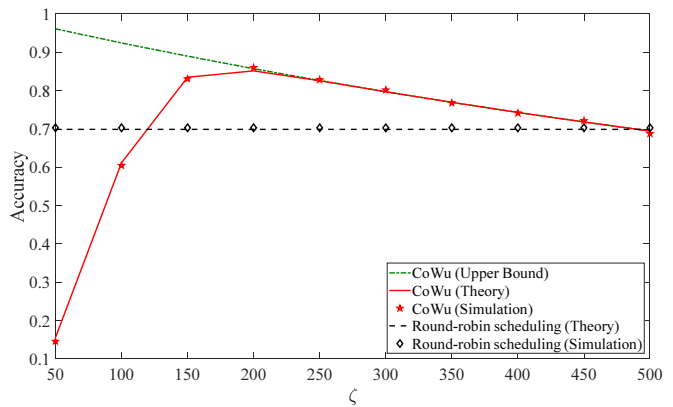}
\caption{Accuracy of CoWu against $\zeta$.}
\label{Fig:zeta_Change}
\end{figure}

Fig. \ref{Fig:zeta_Change} shows the accuracy of CoWu against $\zeta$, where we set $N = 100$, $M = 100$, $[V_{L}, V_{U}] = [94, 98]$, $L = 10$ and $q = 0.0002$. We also plot the upper bound of CoWu, where we set $P_{s}(w_{s},\zeta)=1$ for all $\zeta$ if $w_{s}=w$, otherwise $P_{s}(w_{s},\zeta)=0$, which corresponds to the case where all active users succeed in the communication, and the results of round-robin scheduling, whose accuracy does not depend on $\zeta$. We obtained the numerical results from the theoretical analysis presented in Sec.~\ref{sec:Theoretical_Analysis} and a Monte Carlo simulation over $10^{4}$ transmission rounds. From this figure, we can see that the results
obtained with theoretical analysis coincide with simulation
results, which validates our analysis.

From this figure, we also see an optimal value in terms of accuracy. Let us denote the optimal value of $\zeta$ as $\zeta_{opt}$. For $\zeta < \zeta_{opt}$, we can see that the accuracy becomes smaller as $\zeta$ decreases, because the number of nodes that fail their data transmission by the deadline $T$ increases. For $\zeta > \zeta_{opt}$, most users complete their transmission, but the sensed values become obsolete at the deadline, also leading to a decreased accuracy. We also see that CoWu achieves a higher accuracy than round-robin scheduling, provided that $\zeta$ is selected appropriately. This result illustrates the importance of the timing of the wake-up signaling so as to maximize the accuracy at the deadline. Finally, we can see that CoWu approaches the upper bound as $\zeta$ becomes larger, which is because the error is dominated by obsolete values at the deadline as opposed to collisions.

\subsubsection{Accuracy against $q$}
\begin{figure}[t]
\centering
\includegraphics[width=0.48\textwidth]{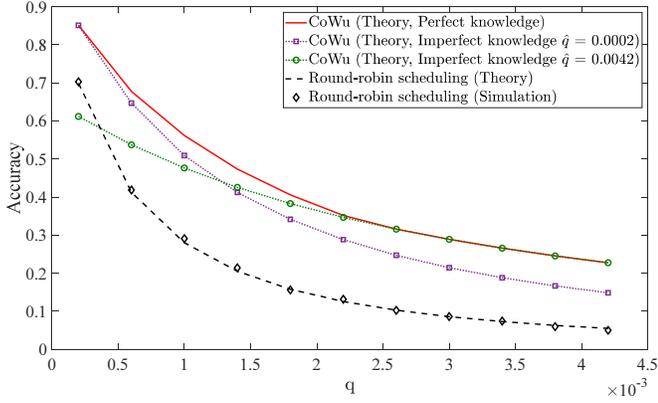}
\caption{Accuracy of CoWu and round-robin scheduling against $q$.}
\label{Fig:results_q_Change}
\end{figure}

Next, we compare the accuracy of CoWu and round-robin scheduling against the speed of the physical process, characterized by $q$. Furthermore, we study the impact of imperfect knowledge about $q$. To this end, we denote the assumed value of $q$ by $\hat{q}$, and use $\hat{q}$ to optimize $\zeta$ in the CoWu scheme. Fig. \ref{Fig:results_q_Change} shows the accuracy of CoWu and round-robin scheduling against the true value of $q$ for the same parameters as in Sec. \ref{sec:D_opt}. 
From the figure, we see that the accuracy of both schemes decreases as $q$ increases, as the data becomes obsolete at the deadline. Next, we see that CoWu can realize higher accuracy than round-robin scheduling across the entire considered range of $q$. In round-robin scheduling, to realize reliable data transmission from all nodes by the deadline, some nodes need to transmit data very early, and their measurements become obsolete at the deadline. This penalty increases as $q$ gets larger. On the other hand, in CoWu only the subset of the nodes that observe values in the requested range at $t_{s}$ wake up and transmit data. With the optimized transmission timing for the wake-up signal, each node can complete data collection by the deadline and convey the timely data toward the sink, thereby obtaining a higher accuracy than with round-robin scheduling.

Finally, we focus on the results of imperfect knowledge. When $\hat{q}=0.2\times 10^{-3}$, i.e., the assumed value is lower than the true $q$, we see that the accuracy deteriorates compared to the one with perfect knowledge as $q$ increases, and thus the difference between the assumed and true $q$ becomes larger. The primary reason for this is that for small $\hat{q}$, the sink chooses a relatively large $\zeta$, because the measurements are unlikely to change before the deadline, and it is more important to ensure that more sensor nodes successfully transmit their measurements. However, because the true $q$ is larger than $\hat{q}$, the data collected at the deadline is likely to be out of date, leading to poor accuracy. When $\hat{q}=4.2\times 10^{-3}$, i.e., larger than the true $q$, we see that the accuracy is deteriorated when $q$ is small. 
In that case, the sink sets a small $\zeta$ to ensure that the received data is timely at the deadline at the cost of a relatively high probability of transmission failures. This leads to an accuracy that is smaller than in the case with perfect knowledge of $q$. 
Note that, despite this deterioration for the cases with imperfect knowledge, the accuracy of CoWu is still higher than that of round-robin scheduling for most cases.

\subsubsection{Energy consumption of nodes}
As mentioned, the energy consumption of the nodes is a critical parameter in IIoT. Here, we compare the total energy consumption of CoWu and round-robin scheduling by Monte Carlo simulation employing the same parameters as in Sec. \ref{sec:D_opt}. 
The power consumption of the transmitting and receiving state is set to 55 mW and 50 mW, respectively \cite{tamura2019low}, and the duration of each time slot is defined as $320~\mu\text{s}$ \cite{TGCN_Content}. 
We ignore the energy consumed by the wake-up receiver, as its power consumption is normally much smaller than that of the main radio.
Here, the total energy consumption is defined as the total amount of energy consumed by sensor nodes during the data collection period. We assume that the node activated by a wake-up signal continues data transmission until it succeeds in data transmission. Thus, the value of total energy consumption only depends on the [$V_{L}, V_{U}$], not on the value of $\zeta$. 
The numerical results show that the average total energy consumption of round-robin scheduling is $17.6~\text{mJ}$, while that of CoWu is $4.50~\text{mJ}$. CoWu can then reduce energy consumption by about $75\%$ against round-robin scheduling: this is achieved by the selective activation of the specific nodes that have informative data, i.e., those in the interval $[V_{L}, V_{U}]$, while all nodes are activated in round-robin scheduling regardless of the importance of their data.

\section{Conclusions}\label{sec:conc}
In this letter, we have applied the CoWu scheme in a scenario in which the timeliness of data at the query timing is crucial. In order to investigate the performance of CoWu, we analytically derived the accuracy for both CoWu and round-robin scheduling, by assuming $p$-persistent CSMA as a MAC protocol. 
Taking advantage of the statistical knowledge about the information on the physical process, we have confirmed that the sink can maximize the accuracy of the query by transmitting a wake-up signal at an adequate timing, and also the superiority of applying CoWu in terms of accuracy and total energy consumption against round-robin scheduling. We have also investigated the case where the statistical prior of the sink is imperfect, and checked the robustness of the schemes to this estimation error. 
Our numerical results showed that CoWu achieves higher accuracy than round-robin scheduling, while reducing total energy consumption by about $75\%$, even in cases with imperfect estimation.

Possible avenues for future work include the investigation of a general case where the query arrives periodically, and a comparison between CoWu and the scheduling method considering a more realistic channel model and imperfect knowledge of the physical process is also an interesting direction.

\ifCLASSOPTIONcaptionsoff
  \newpage
\fi

\bibliographystyle{IEEEtran}
\bibliography{IEEEabrv,main}
\end{document}